# Morphology and magneto-transport in exfoliated graphene on ultrathin crystalline β-Si$_3$N$_4$(0001)/Si(111)

Sedighe Salimian,[1] Shaohua Xiang,[1] Stefano Colonna,[2] Fabio Ronci,[2] Marco Fosca,[2] Francesco Rossella,[1] Fabio Beltram,[1] Roberto Flammini[2,*], and Stefan Heun[1,†]

[1] NEST, Istituto Nanoscienze-CNR and Scuola Normale Superiore, Piazza San Silvestro 12, 56127 Pisa, Italy

[2] CNR-ISM Istituto di Struttura della Materia, Via del Fosso del Cavaliere 100, 00133 Roma, Italy

We report the first experimental study of graphene transferred on β-Si$_3$N$_4$(0001)/Si(111). Our work provides a comprehensive quantitative understanding of the physics of ultrathin Si$_3$N$_4$ as a gate dielectric for graphene-based devices. The Si$_3$N$_4$ film was grown on Si(111) under ultra-high vacuum (UHV) conditions and investigated by scanning tunneling microscopy (STM). Subsequently, a graphene flake was deposited on top of it by a polymer–based transfer technique, and a Hall bar device was fabricated from the graphene flake. STM was employed again to study the graphene flake under UHV conditions after device fabrication and showed that surface quality is preserved. Electrical transport measurements, carried out at low temperature in magnetic field, revealed back gate modulation of carrier type and density in the graphene channel and showed the occurrence of weak localization. Under these experimental conditions, no leakage current between back gate and graphene channel was detected.

[*] e-mail: roberto.flammini@cnr.it
[†] e-mail: stefan.heun@nano.cnr.it

# 1. INTRODUCTION

In the last decade, graphene and graphene-derived systems have attracted wide attention in view of possible applications in electronics and sensing that are envisioned in light of graphene's unique electrical, mechanical, thermal, and optical properties combined with its intrinsic 2D nature [1-4]. Experimentally-accessible transport properties of graphene on standard $SiO_2$ substrates are however limited by roughness, charge scattering, and impurities [5-7]. Alternative dielectric substrates with small lattice mismatch with graphene and possibly with a high-k dielectric constant are therefore desirable for improving graphene-based device performance. Hexagonal boron nitride (h-BN) is appealing for this purpose owing to its relatively smooth surface virtually free of charge traps. Indeed, graphene devices on exfoliated h-BN substrates have higher mobilities than those fabricated on $SiO_2$ [8], however, their technological relevance is limited by the small size of the exfoliated h-BN flakes. On the other hand, large-scale polycrystalline h-BN grown by state-of-the-art techniques requires to be transferred to target substrates [9,10]. Among high-k dielectric materials, $Si_3N_4$ ($\varepsilon \approx 6.6$) [11] is regarded as one of the best candidates for this application. An ultrathin layer (less than 1 nm) of wafer-scale crystalline $\beta$-$Si_3N_4$(0001) can be readily grown on Si (111) substrates [12] and is known to passivate the Si (111) surface [13,14]. Indeed, studies on metal/$\beta$-$Si_3N_4$(0001)/Si(111) interfaces showed that no reaction takes place between metal (Au [13,15], Co [16,17], Fe [18]) and silicon when the nitride interlayer is present. Besides, the surface of $\beta$-$Si_3N_4$(0001) is only mildly affected by air exposure, and the nitride surface preserves its electronic properties (and that of the silicon substrate beneath) upon thermal oxidation [19].

The properties of $\beta$-$Si_3N_4$ as a substrate for graphene-based electronics were theoretically investigated by Yang et al. [20]. The authors demonstrated that the interaction between graphene and $\beta$-$Si_3N_4$ is lower than with SiC, or with any oxide substrate that typically induce graphene-surface deformation thus degrading its transport properties. Additionally, $\beta$-$Si_3N_4$ produces the lowest mechanical strain on graphene thanks to the small lattice mismatch [20]. These characteristics are predicted to decrease the impact on electron mobility by reducing the inhomogeneity of charge distribution in graphene. In fact, Yang et al. predicted the possibility to achieve ultra-high electron mobility in graphene supported by $\beta$-$Si_3N_4$, which therefore can be a promising building block for the future of graphene technology.

Examples of graphene/silicon nitride interfaces, mainly in view of device fabrication, were already reported [11,21-24], but the methods employed for interface fabrication relied on nitride growth on top of the graphene layer [11,22] or on high-temperature graphene growth on top of nitride surfaces [23,24]. In both cases, the resulting silicon nitride layer is amorphous or polycrystalline and must be thick, in order to preserve its insulating properties. At variance, $\beta$-$Si_3N_4$ films on Si(111) are known to have a crystalline structure [25,26] and can be very thin [27]. Although the dielectric properties of $Si_3N_4$ may undergo a change when reducing the film thickness to the nanoscale, literature reports that such films still keep their bulk fundamental band gap of about 5 eV [28].

In this paper, we used a polymer-assisted method [29-31] for the transfer of exfoliated graphene onto crystalline $\beta$-$Si_3N_4$, an approach that is reliable and compatible with surface-science techniques, known to be highly demanding in terms of cleanness and atomic order. After an initial preparation and characterization of the nitride surface at the atomic level under UHV conditions, a graphene flake was transferred on top of it. Micro–Raman spectroscopy was employed to verify graphene quality, and electron-beam lithography followed by metal evaporation was used to fabricate Ohmic contacts on the graphene layer in a Hall-bar configuration. The graphene surface was subsequently characterized by STM under UHV conditions, indicating that even after complete-device fabrication the structural properties of the graphene are preserved. Back-gate modulation on the graphene channel, weak localization, and conduction through the dielectric were investigated by magneto-transport measurements at 4.2 K.

## 2. EXPERIMENTAL

Si(111) substrates were cut from p-type (B-doped) silicon wafers with resistivity $\rho \cong 0.005$ $\Omega$cm. The surface was outgassed at 700 K for several hours using direct resistive heating. The Si substrate was cleaned by repeated flashes of several seconds at 1520 K at a base pressure of $\sim 3\times10^{-10}$ mbar, leading to a sharp (7 × 7) low energy electron diffraction (LEED) pattern. The (7 × 7)-reconstructed surface was held at a temperature of approximately 1050 K and exposed to 100 Langmuir of $NH_3$. This yields a sample surface evenly covered by $\beta$-$Si_3N_4$ [25-27,32,33].

The STM measurements were carried out in constant-current mode at room temperature using an Omicron LT-STM housed in an ultra-high vacuum (UHV) chamber ($5 \times 10^{-11}$ mbar base pressure). Electrochemically-etched tungsten tips were used after a cleaning procedure by electron bombardment. The reported bias voltage ($V_S$) is referred to the sample.

Graphene flakes were transferred on $\beta$-$Si_3N_4$(0001)/Si(111) substrates by a PMMA/PVA-assisted transfer method [31]. The resulting graphene/$Si_3N_4$ samples were annealed at about 420 K for 4 hours before UHV-STM surface investigation.

Raman spectra were acquired with a Renishaw inVia system using a 532 nm excitation wavelength. Electrical transport and magnetoresistance measurements were performed in a Heliox cryostat. Two-probe measurements were obtained with a Keithley 2614B power supply, while four-probe measurements were carried out by lock-in technique with a 100 nA current injected in the graphene channel.

## 3. RESULTS

### 3.1. The $\beta$-$Si_3N_4$(0001)/Si(111) substrate

Figure 1(a) shows an STM image of the silicon nitride surface. The image shows the typical appearance of a sample upon thermal nitridation [32,34]. The surface is characterized by large terraces with an average size of about 100 nm in diameter. In Fig. 1(b), the numerical derivative along the horizontal direction of a region of Fig. 1(a) is shown. The image reveals that even between islands a nitride layer is present. The result of the nitridation is a conformal film of 1 nm in thickness [27,32] without apparent cracks. From Fig. 1(a), a roughness value of 0.503 nm is inferred, resulting from the substrate morphology.

This is consistent with previous results that demonstrate the absence of any silicide formation when a reactive metal is deposited on top of the nitride surface [13,15-18]. Figure 1(c) reports the LEED pattern of the sample: an (8×8) reconstruction is observed thus demonstrating that the film is a single $\beta$-$Si_3N_4$ crystal [25,26]. In panel (d) we report an atomically-resolved image from the same sample that shows high surface ordering. The surface unit cell [26] is highlighted by a blue rhombus.

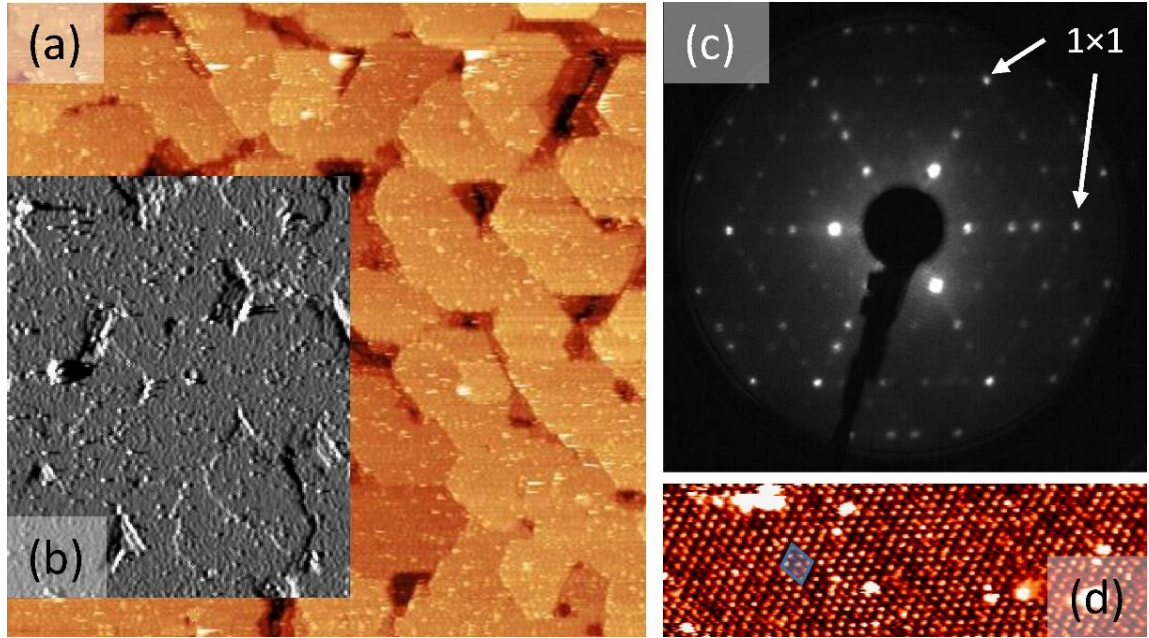

FIG. 1. (a) STM image of the β–Si$_3$N$_4$ (0001) surface (500×500 nm$^2$, +4 V, 0.5 nA), (b) derivative along the x-direction of part of the image in (a), (c) LEED pattern showing an (8×8) reconstruction (E = 52.6 eV), (d) atomically resolved STM image of the nitride surface (50×16 nm$^2$, -4V, 0.1 nA).

### 3.2. Graphene on β–Si$_3$N$_4$(0001)/Si(111): Micro–Raman spectroscopy

Graphene transfer on the target β-Si$_3$N$_4$(0001)/Si(111) substrate was carried out using the PMMA/PVA assisted transfer method [31]. An optical-microscopy image of the exfoliated graphene flake on PVA/PMMA is shown in Fig. 2(a). Two graphene flakes show different contrast in the optical microscope: on the left, a monolayer flake is indicated by the white arrow, while on the right, top and bottom parts of the flake feature a double and triple graphene layer, respectively. In this case we selected the monolayer flake for transfer and device processing.

Graphene flakes were then transferred on the nitride surface and the PMMA removed. At this point, the graphene flake is ready for contact fabrication. However, graphene is not visible under the optical microscope unless a suitable thickness of the insulating layer along with its correspondent refractive index is used. Systematic studies found that the visibility of graphene on thin silicon nitride films is very low [35]. Here, the ultrathin dielectric did not give sufficient contrast to visualize graphene by optical imaging, and therefore a Raman map was used to determine graphene position and the boundaries for device fabrication.

Furthermore, we used Raman spectroscopy to evaluate the quality of the transferred graphene flake. Figure 2(b) shows two spectra taken on the nitride surface and on graphene. No silicon-nitride Raman-active modes are present in this energy range [36], therefore no peaks are detected on the nitride surface, as expected. Analogously no peaks related to physisorbed species, namely N$_2$ and O$_2$, at about 1550 and 2300 cm$^{-1}$ [36], respectively, are present. On the contrary, the expected G and 2D peaks are observed on the graphene flake. They are attributed to Raman scattering with one-phonon emission and to the double resonant Raman scattering with two phonon emission, respectively [37]. The small peak at about 2450 cm$^{-1}$ is generally attributed to two-phonon modes [38]. The integrated intensity ratio of the 2D peak (∼2700 cm$^{-1}$) to that of the G peak (∼1580 cm$^{-1}$) indicates the presence of monolayer graphene, while the absence of the D band (∼1350 cm$^{-1}$) is consistent with the high-quality of exfoliated graphene [37,39–41].

Micro-Raman maps of the flake were recorded by selecting short bandwidths centered at the two main peaks of Fig. 2(b). These maps are shown in Fig. 2(c-e). The flake shape is well reproduced by tracking the 2D or G Raman modes. On the other hand, in the map measured when the D peak is selected, the flake is not visible, confirming the high quality of the graphene flake.

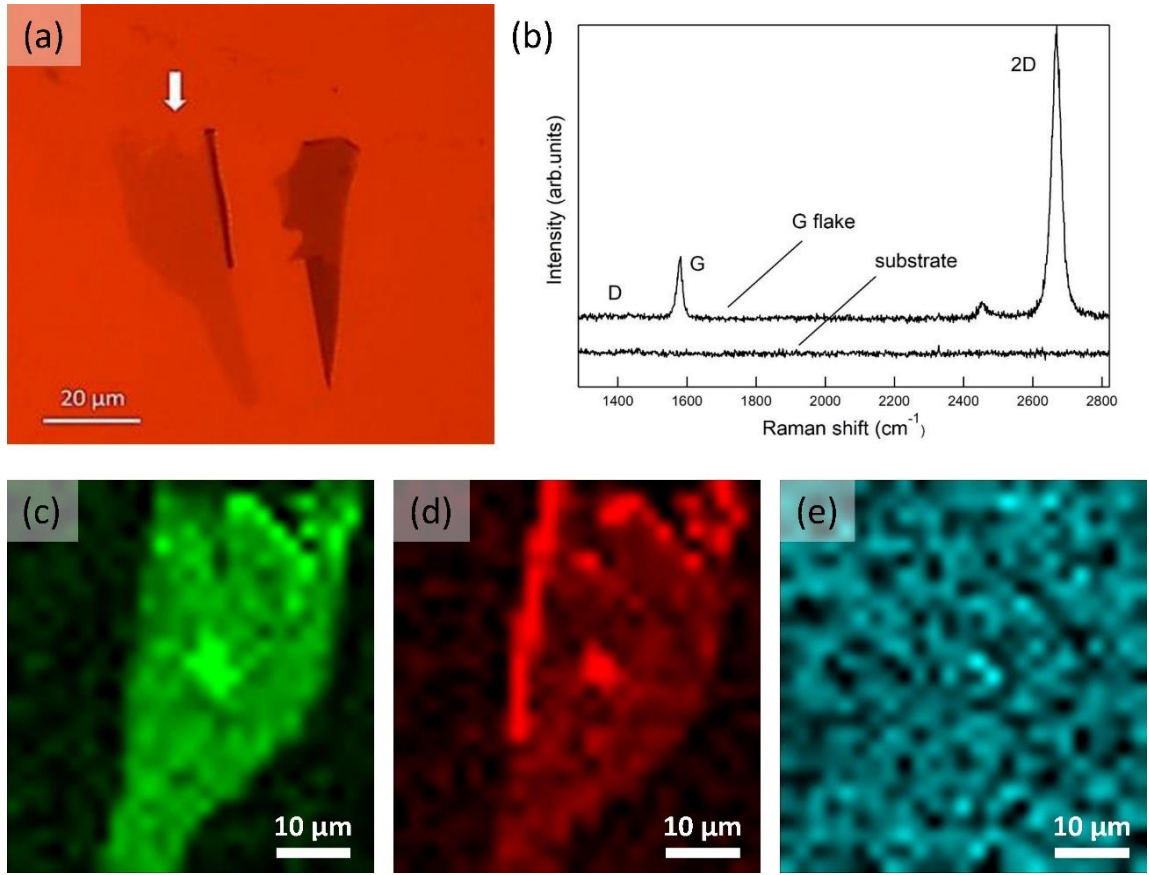

FIG. 2. (a) Optical microscopy image of two graphene flakes on PMMA. The arrow shows the monolayer graphene flake to be transferred on top of the nitride surface. (b) Raman spectra taken on the graphene flake and on the nitride surface. The high quality of the graphene flake after transfer is demonstrated by the absence of the D peak, generally ascribed to defects. No peaks are expected from the silicon nitride layer in this energy range. (c-e) Micro–Raman maps of the graphene flake taken in correspondence with the 2D, G and D peaks of Fig. 2(b), in panels (c), (d) and (e), respectively.

### 3.3. STM on graphene/β–Si$_3$N$_4$ after device fabrication

It is well known that a PMMA-based transfer processes can affect the surface quality of graphene in devices [42]. This prompted a careful check of the graphene surface condition after device fabrication. To this end, we complemented the Raman spectroscopy data with atomic structure information. This was done by STM, working under UHV conditions. Figure 3(a) shows a 10×10 nm$^2$ image of the graphene flake, demonstrating its high quality. In panel (b) we show a magnification of the graphene surface with atomic resolution (4×4 nm$^2$). The background was subtracted and a profile traced (along the blue line in panel (b)): see panel (d). The image shows a lattice period of 0.256 nm, in agreement with the literature [43,44]. The image shows a triangular lattice instead of the typical honeycomb lattice structure characteristic of non-interacting graphene. Based on the Raman data, we can exclude that this is due to the presence of bi- or even tri-layer graphene. At variance, we attribute this behavior to the effect of a slight curvature [5,45], as shown by the larger size image of panel (a).

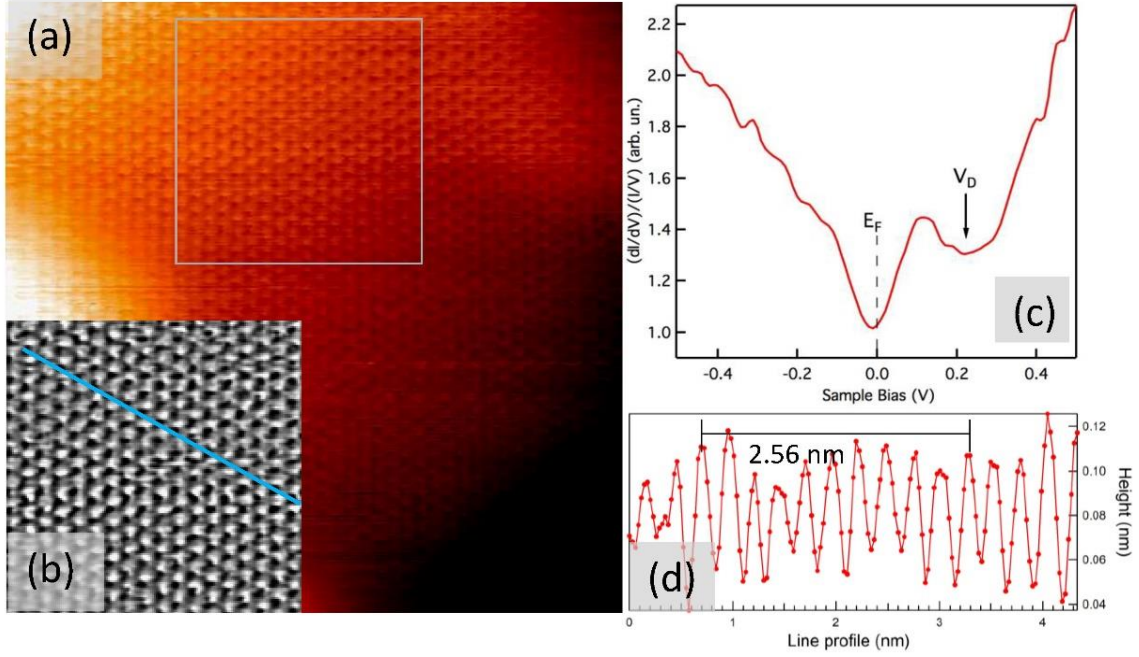

FIG. 3. (a) 10 × 10 nm$^2$ STM image of the graphene flake after device fabrication. Imaging parameters: -0.5 V, 5 nA. In (b), a magnification (4×4 nm$^2$) of the highlighted square area in panel (a) shows the fine structure of the graphene flake. (c) STS spectrum measured on the same area as panel (a). The spectrum is the result of an average made from several measurements (see text). All spectra were collected with a lock-in amplifier ($I_t$ = 5 nA). A line profile of panel (b) is shown in (d). It shows that 10 periods correspond to 2.56 nm.

The local electronic properties of the graphene flake were investigated by scanning tunneling spectroscopy (STS) measurements. The STS spectrum reported in Fig. 3(c) was obtained by averaging the spectra collected on a grid evenly distributed on the surface of the STM image of panel (a), with a tunneling current set point of 5 nA. The spectrum shows a gap-like feature at around 0 V (i.e. at the Fermi level) and a dip at +0.21 V. The gap-like feature was already observed in STS measurements of graphene [46-48]. This feature is attributed to the suppression of electronic tunneling to graphene states with a large wave vector near the Fermi energy and to the enhancement of electronic tunneling at higher energy owing to a phonon-mediated inelastic channel. In more detail, tunneling from a tip (with rotational symmetry) typically occurs at the Γ point (at the center of the Brillouin zone). The peculiar band structure of graphene shows Dirac cones at the K/K' points (at zone boundary), but a band gap at the Γ point. Since no states are available in the elastic channel (around the Γ point), electrons can only tunnel into graphene if they exchange momentum with lattice phonons. Such inelastic process is only possible if the bias voltage is larger than the phonon energy. For graphene, the out-of-plane acoustic phonon mode has an energy of 67 meV, consistent with the half width at half maximum of the gap-like feature.

The STS measurement in Fig. 3(c) also shows a dip located around +0.21 V, indicated as $V_D$. This feature is attributed to the energy position of the Dirac point [46-48]. Subtracting the energy of the out-of-plane acoustic graphene phonon mode (67 mV), the Dirac point is more precisely located at 0.14 eV above the Fermi energy, indicating p-type doping in the graphene with a hole concentration of $1.4\times10^{12}$ cm$^{-2}$.

From theoretical calculations of the graphene/Si$_3$N$_4$ system, a slight n-type doping of the graphene in contact with the nitride substrate would be expected [20]. This discrepancy can be explained by considering that the graphene transfer process is performed in air. It is likely that, as demonstrated for similar systems [50-52], water molecules may be adsorbed on the nitride surface (being the process exothermic [49]) and trapped at the interface with graphene. It is important to note that the mild thermal treatment of the graphene prior to the STM measurement can assure contaminant desorption from the sample surface, but water may remain trapped at the interface even at rather high temperature, as demonstrated in the case of graphene on mica [52]. This water layer has a twofold effect on the graphene layer. On the one hand, it is able to screen the electronic interaction with the substrate [52], thus hindering the predicted n-type doping from the nitride. On the other hand, it can induce a p-type doping of the graphene due to the polar nature of the water molecule, as demonstrated for other substrates [50,51]. This explains also why no moiré structure was recorded in the STM images: the interaction between graphene and nitride is probably mediated by a thin water layer that decouples the graphene layer from the substrate, as suggested by the Raman spectra and typical of non-interacting graphene.

**3.4. Magneto-transport measurements**

Device fabrication was initiated by Hall-bar design followed by a step of e-beam lithography (EBL), 10/100 nm Cr/Au deposition for the contacts, and lift-off. Figure 4(a) shows a sketch of the Hall-bar device. Metal leads on top of the nitride surface are made possible by nitride's ability to prevent the formation of silicide. In order to avoid any adverse effect of wire bonding on the ultrathin Si$_3$N$_4$ underneath the gold pads, a second step of EBL was performed. Lithographic holes are patterned on top of the gold pads and then filled with 135 nm SiO$_2$ followed by lift off. In another step of lithography, 10/150 nm Cr/Au is deposited on top of the SiO$_2$. The steps between large area SiO$_2$ and pre-patterned electrodes are patched in order to achieve continuous metal contacts.

The top of Fig. 4(a) shows an optical-microscopy image of the Hall-bar device. The distance between source and drain is $L_{sd} = 18$ µm, and the length $L$ between the Hall-bar contacts 1-2 and 3-4 is $L = 10$ µm, while the width $W$ between the contacts 1-4 and 2-3 is $W = 18$ µm. The graphene flake is localized between source and drain, but is not visible in optical microscopy: a dashed line in the inset indicates the boundary of the graphene flake as obtained from micro-Raman maps.

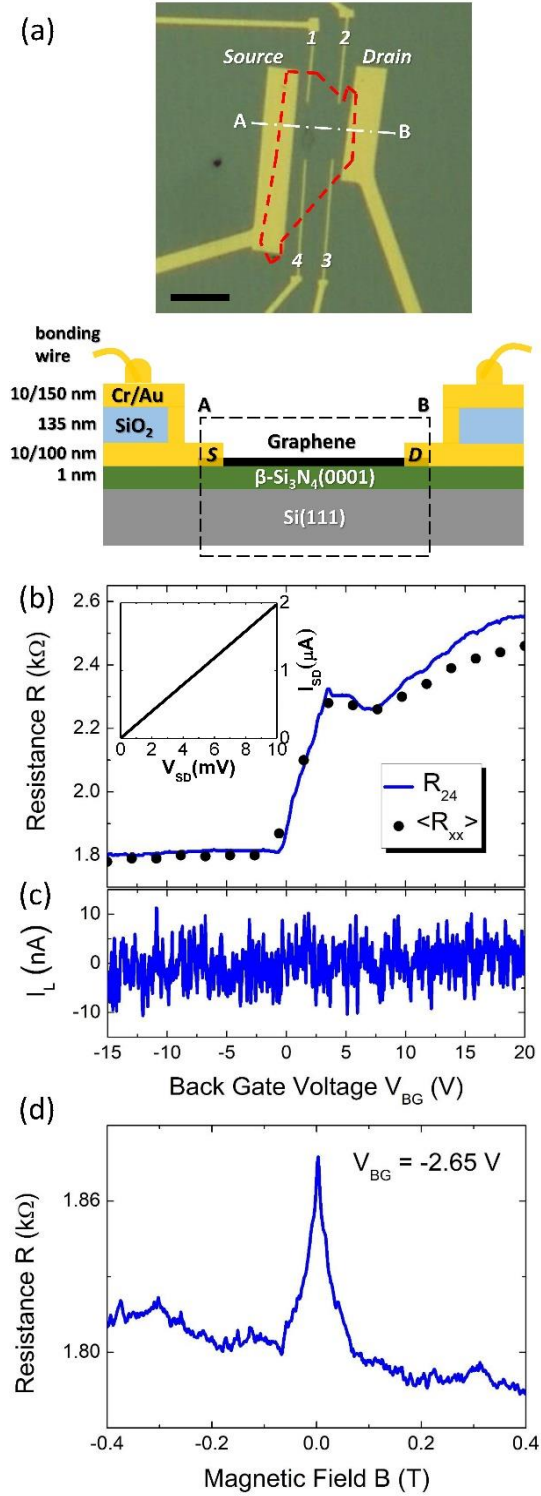

FIG. 4. (a) Top: Optical microscopy image of the Hall bar device. Yellow areas indicate the metal electrodes. The dashed red line indicates the border of the graphene flake (not visible by optical microscopy). Scale bar 20 µm. Bottom: Sketch of the device, showing a cross-section along line A-B indicated in the optical microscopy image. (b) Channel resistance $R$ as a function of back gate voltage $V_{BG}$. Both the electrical transport curve ($R_{24}$, solid line) and the average $R_{xx}$ values obtained from the Hall measurements at different back gate voltages (dot symbols) are shown. Inset: Current-voltage ($I_{SD}$-$V_{SD}$) curve measured between source and drain contacts at 4.2 K in two-probe configuration (at back gate voltage $V_{BG} = 0$ V). $R = 5$ kΩ. (c) Loss current $I_L$ from the back gate to the sample as a function of back gate voltage, measured simultaneously with the $R_{24}$ data shown in (a). (d) Resistance $R$ between contacts 2 and 4 as a function of magnetic field $B$ at back gate voltage $V_{BG} = -2.65$ V.

Current-voltage ($I_{SD}$-$V_{SD}$) curves measured at 4.2 K in two-probe configuration (at back gate voltage $V_{BG}$ = 0 V) are linear indicating the presence of good Ohmic contacts between graphene and the metal electrodes, see the inset to Fig. 4(b). In a four-probe measurement, a constant current (100 nA) was injected between source and drain, while the voltage drop was measured between contacts 2 and 4 of the Hall bar. Figure 4(b) shows the gate-voltage dependence of the resulting graphene resistance $R_{24}$, measured at 4.2 K. Starting from a back-gate voltage of about 0 V the resistance strongly increases with increasing back-gate voltage, reaching a local maximum at about 3.5 V, which suggests that the charge neutrality point is located there. This indicates that the graphene is p-doped, consistently with the STM data. The $R_{24}$ vs. $V_{BG}$ data shown in Fig. 4(b) was measured up to a back-gate voltage of 20 V, a range in which no leakage current from the gate to the graphene could be detected (see Fig. 4(c)).

Figure 4(b) also shows the presence of a shoulder for higher positive back-gate voltages, for which the resistance does not return to the values measured for negative back gate voltages. This could be due to disorder [53] or inhomogeneities in the graphene channel [54], or it could stem from the presence of a satellite Dirac point [55]. In principle such satellite Dirac point could be induced by van der Waals interaction between graphene and a crystalline substrate that can lead to the formation of a moiré pattern, resulting in a long-wavelength superlattice potential in graphene. This was predicted [56] and observed [55] to lead to the formation of additional satellite Dirac points for graphene on hBN. Likewise, the surface lattice constant of β-$Si_3N_4$(0001) matches well with that of the 3x3 graphene primitive cell [20], and therefore the observation of additional Dirac points in a low-angle-aligned $Si_3N_4$-graphene heterostructure could be expected. However, since no moiré pattern was observed by STM in this work, we tend to exclude this as the reason for the appearance of a shoulder. Rather, we believe that this more complex electron dispersion is caused by metal-induced doping in the contact regions (without a pinning of the work function there). In fact, similar effects were already observed for graphene on $SiO_2$ [57,58] and on $SiN_x$ [21].

Carrier concentration and mobility were obtained through Hall-effect measurements. We performed a series of magneto-transport measurements at different back-gate voltages at 4.2 K. An example is shown in Fig. 4(d) for $V_{BG}$ = -2.65 V where we plot a magnetoresistance measurement between contacts 2 and 4: owing to the diagonal geometry of the measurement, data include a mixed longitudinal and transverse voltage drop. Note also the peak at 0 T that we attribute to weak localization [59,60]. We analyzed such curves taking into account the odd and even components in the trace in order to extract the carrier concentration and mobility (see SI for the detailed procedure adopted). The negative sign of the slope in Fig. 4(d) indicates p-type behavior of the graphene channel, in agreement with the data shown in Fig. 4(b). Black dots in Fig. 5(a) provide the measured experimental hole-concentration values as a function of $V_{BG}$.

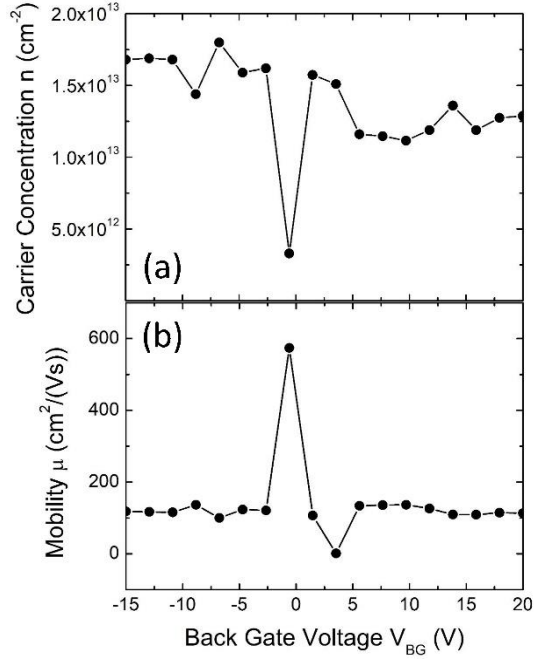

FIG. 5. (a) Charge carrier concentration $n$ and (b) mobility $\mu$, evaluated from the Hall effect measurements.

Apart from the weak-localization peak, $R_{xx}$ is weakly dependent on magnetic field for small intensities and we calculated the average values of $R_{xx}$ including the weak localization peak (in the range $|B| < 1$ T). The average values of $R_{xx}$ are plotted in Fig. 4(b) as dot symbols for the different back-gate voltages. The trend of $\langle R_{xx}\rangle$ is fully consistent with the behavior of the field-effect modulation curve (solid line) and confirms the validity of our analysis.

Using the measured carrier concentrations (Fig. 5(a)) and the average values of $R_{xx}$ (Fig. 4(b)), we can determine the graphene Hall mobility $\mu$ for the different back-gate voltages:

$$\mu = \frac{L}{W\langle R_{xx}\rangle} \frac{1}{ne} \quad . \quad (5)$$

The resulting carrier mobilities are shown as dot symbols in Fig. 5(b).

While the leakage current between back gate and device, across the $Si_3N_4$ dielectric, at 4.2 K remained below the noise floor of the voltage supply (~10 nA, see Fig. 4(c)), at room temperature we detected a sizeable leakage current that was as high as 0.4 µA under a back-gate bias of 1 V. Therefore, no back-gate modulation could be measured at room temperature. Figure 6 shows the values of leakage current as a function of temperature upon increasing device temperature from 4.2 K to 300 K, measured with an applied back gate voltage of 1 V. Up to approximately 20 K, the leakage current is below the detection limit of the power supply (~10 nA, see the inset to Fig. 6). It then increases strongly with increasing temperature up to 50 K, and then saturates to an approximately constant value. This rapid increase in the leakage current at 20 K is not consistent with the relevant metal-insulator-semiconductor transport mechanisms (tunnel, thermionic or Poole-Frenkel) [61] and we believe can be linked to weak spots in the dielectric likely caused by the fabrication steps.

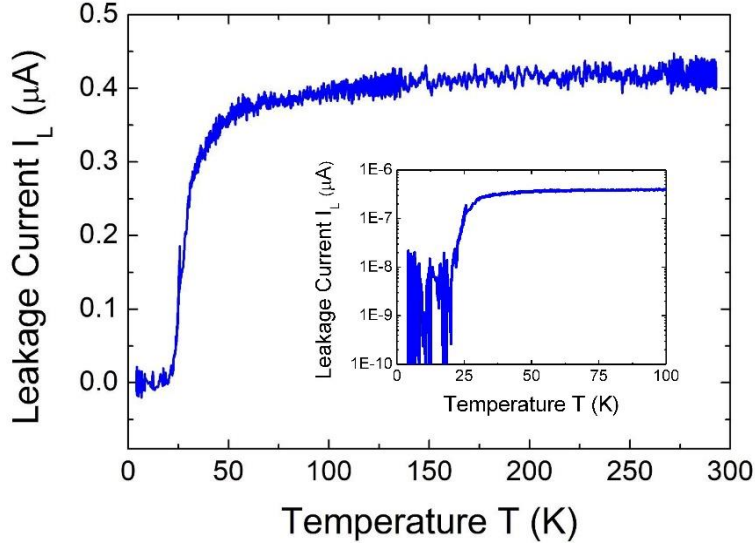

FIG. 6. Leakage current $I_L$ as a function of temperature at constant $V_{BG} = 1$ V. The inset shows the same data (up to 100 K) on a log-scale, to highlight the noise floor of the measurement.

## 4. CONCLUSIONS

We reported on the suitability of β-Si$_3$N$_4$(0001) as a substrate for the fabrication of high-quality graphene devices. For this purpose, we studied thin (< 1 nm) β-Si$_3$N$_4$(0001)/Si(111)-(8×8) samples by STM. Graphene flakes were deposited on the nitride film by a PMMA-based transfer technique. A Raman study was performed in order to ascertain the position of the graphene flake and its quality. Gold leads and contacts were lithographed on the surface in order to fabricate a Hall-bar device. The resulting device was then studied again by STM in UHV. STM images showed the high quality of the graphene in the completed device. Magneto-transport results demonstrated that crystalline β-Si$_3$N$_4$ (0001) is a good support for graphene because of its high-κ properties, the low lattice mismatch with graphene, and the low surface roughness. Magneto-transport measurements confirmed the p-type doping of the graphene flake measured by STS.

## ACKNOWLEDGEMENTS

We thank Filippo Fabbri for his help with the elaboration of the Raman data.